\documentstyle[titlepage,12pt]{article}
\begin{document}
\renewcommand{\theequation}{\thesection.\arabic{equation}}
\title{Quantum Coherent String States in $AdS_3$ and $SL(2,R)$ WZWN Model}
\author{ A.L. Larsen\thanks{Department of
Physics, University of Odense,
Campusvej 55, 5230 Odense M, Denmark.}
and
N. S\'{a}nchez\thanks{Observatoire de Paris,
DEMIRM. Laboratoire Associ\'{e} au CNRS
UA 336, Observatoire de Paris et
\'{E}cole Normale Sup\'{e}rieure. 61, Avenue
de l'Observatoire, 75014 Paris, France.}}
\maketitle
\begin{abstract}
In this paper we
make the connection between semi-classical string quantization
and exact conformal field theory quantization of strings in 2+1 Anti de Sitter
spacetime. More precisely, considering
the WZWN model corresponding to $SL(2,R)$ and its covering group,
we construct quantum
{\it coherent} string states, which generalize the ordinary coherent
states of quantum mechanics,  and show that in the classical limit
they correspond to
oscillating
circular strings. After quantization, the spectrum is found to consist of two
parts: A continuous spectrum of low mass states (partly tachyonic) fulfilling
the standard spin-level condition necessary for unitarity $|j|< k/2$,
and a discrete spectrum
of high mass states with asymptotic behaviour $m^2\alpha'\propto N^2$
($N$ positive
integer).
The quantization condition for the high mass states arises from the
condition of finite positive norm of the coherent string states, and the result
agrees  with our previous results obtained using semi-classical
quantization.
In the $k\rightarrow\infty$ limit, all the usual properties of coherent or
{\it quasi-classical} states are recovered.

It should be stressed that we consider the circular strings
only for simplicity and clarity, and that our construction can easily be used
for other string configurations too.
We also compare our results with those obtained in the recent
preprint hep-th/0001053 by Maldacena and Ooguri.
\end{abstract}
\newpage
\section{Introduction}
\setcounter{equation}{0}
The question about
the spectrum of bosonic string theory in 3-dimensional Anti de Sitter space,
$AdS_3\cong\; SL(2,R)\cong\; SU(1,1) $,
attracted a lot of interest about 10 years ago [1-10], as a first example
of exact string quantization on a manifold with curved space {\it and} curved
time. It was immediately realized \cite{balog}
that the problem of unitarity was much more
complicated than in flat Minkowski spacetime \cite{god,brow},
in the sense that the Virasoro
constraints for $AdS_3$ themselves did not eliminate all negative norm states
\cite{balog}.
However, in a series of papers, going back to \cite{dixon,petr,moh}
(see also [6-10, 13, 14]), it has been
argued that unitarity can be ensured (at least for free strings) by imposing
certain restrictions on the allowed representations, exemplified by the now
well-known spin-level restriction for the discrete representations $|j|<k/2$,
where $j$ is the spin and $k$ is the level of the $SL(2,R)$ WZWN model. For
a somewhat different approach towards unitarity, see \cite{ibars,ysatoh}.

More recently, the interest in $AdS_3$ (as well as higher dimensional $AdS$
spaces) has increased in connection with the  conjecture \cite{malda}
relating supergravity and superstring
theory on $AdS$ space with a conformal field
theory on the boundary.
In such constructions, $AdS_3$ often appears on the 10-dimensional
supergravity/superstring
side in a cartesian product with some other compact spaces, for instance
as $AdS_3\times S^3\times T^4$. Thus, again it has become extremely
important to understand the precise spectrum of string theory in $AdS_3$.

Even if the problem of unitarity appearently could be solved by the
spin-level condition (although this question is certainly not completely
settled yet), several problems remained. One of the most important
being that the spin-level restriction together with the mass-shell condition
imposes a restriction on the grade (to avoid confusion with the level $k$, we
use the word "grade" for what is usually called level).
This restriction on the grade means that for fixed level $k$, a string living
in $AdS_3$ can only be excited to its very lowest grades. In other words, we
are faced with the problem that it
seems impossible to have very massive strings in $AdS_3$.

On the other hand, the dynamics of classical strings and their semi-classical
quantization in $AdS_3$ [18-22] does
not seem to indicate any particular problems
for very long and very massive strings, although the question of unitarity
cannot really be addressed exactly in such studies.
It is therefore highly interesting and important to understand how such
long massive strings can arise in an {\it exact} quantization scheme for
strings
in $AdS_3$, without being in conflict with unitarity.

In this paper we suggest that long massive strings can be described as
{\it coherent}
string states based on one of the standard discrete representations
of $SL(2,R)$. For simplicity and clarity, we shall construct quantum
coherent string
states corresponding, in the classical limit, to the
circular strings discussed in
\cite{vega},
but our construction can be used for other string configurations too.

As for (most) other families of string states in $AdS_3$, coherent string
states generally do not have positive norm, even if they fulfil the Virasoro
conditions. The condition of finite positive norm for the coherent states
gives rise to certain restrictions on the spin $j$, which in turn restricts
the mass of the states. We show that the finite positive norm condition for
our coherent string states leads to a mass-spectrum consisting of two parts:
A continuous spectrum of low mass states (partly tachyonic) where $j$
fulfils the standard spin-level restriction, as well as an infinite tower of
discrete high mass states for which the mass-formula is given by eq.(5.20)
and  asymptotically is $m^2\alpha'\propto N^2$
($N$ integer). This result agrees, to leading order, with what was found
using semi-classical quantization \cite{san2,vega1,vega}.

When completing this paper, a recent preprint by Maldacena and Ooguri appeared
\cite{Ooguri}, considering a similar problem. Our construction, however, is
completely different from theirs. First of all, their massive strings are
based on descendents of
primary states for a new set of $SL(2,R)$ representations obtained from the
standard ones by a "spectral flow" operation \cite{maans2},
whereas our massive strings are based on
coherent states of descendents of primary states from
the standard $SL(2,R)$ representations. Secondly, their construction
relies heavily on the existence of some internal compact  manifold
${\cal M}$, which is
assumed to give a large contribution to the total world-sheet energy-momentum
tensor, whereas our construction works directy for $AdS_3$ without need of any
additional internal compact manifold (although of course we could  easily
include an internal compact manifold as well). In
fact, with the internal compact manifold, the
construction of \cite{Ooguri} gives only a finite number of very massive
states, and without the internal manifold it gives
at most a few very massive states (or even none, depending on some other
parameters). Our construction gives in any case an {\it infinite} tower
of more and more massive states in the $AdS_3$ and $SL(2,R)$ background.
This is again in agreement with the previous semi-classical quantization
results \cite{san2,vega1,vega} giving an infinite number of string states
in the $AdS_3$ and $SL(2,R)$ background. Yet another difference between the two
approaches has to do with the world-sheet energy $L_0$. In the construction of
\cite{Ooguri}, $L_0$ is not bounded from below for the representations
obtained by the spectral flow, while in our construction $L_0$ is bounded from
below since we are using the standard representations. However, we are not
working
with eigenstates of $L_0$ thus the standard mass-shell condition
$(L_0-1)|\psi>=0$
is replaced by $<\psi|(L_0-1)|\psi>=0$. In any case, the mass-shell condition
eventually selects those states with "$L_0=1$".

It must be noticed however, that in despite of the differences between
the construction of \cite{Ooguri} and our construction, it turns out that
the final results for the mass-spectrum (at least when an
internal compact manifold with a large contribution to the world-sheet
energy-momentum tensor is assumed) are more or less identical; namely,
a low mass continuous spectrum and a high mass discrete spectrum, where
the energy (or mass) to leading order grows linearly with an integer.

Interestingly enough, our quantum coherent states are a
string generalization of the ordinary coherent states of quantum mechanics.
All the usual properties of ordinary coherent states \cite{merz} are
obtained in the $k\rightarrow\infty$ limit. For instance, the low mass
continuous spectrum of string states become the ordinary coherent states,
eigenstates of the annihilation operator, for any value of the spin
$j\leq -1/2$, while the high mass discrete spectrum of string states completely
disappears, pushed towards infinite mass. These are precisely the properties
which in quantum mechanics characterise coherent states as
{\it quasi-classical},
being the states for which quantum uncertainty is minimal.

Our paper is organized as follows.
In Section 2, we review the classical $SL(2,R)$ WZWN model, mainly to
set our conventions and normalizations. We also give a simple derivation of
the reduction of the classical equations of motion to the Liouville equation
\cite{liou,raif}. In Section 3, we reconsider the classical oscillating
circular strings  \cite{vega} in terms of $SL(2,R)$ currents.
In Section 4, we present the standard results of the quantization of
conformal field theories on a group manifold \cite{wit2}; we only give the
results which we will use later. In Section 5, we then turn
to the construction of the quantum
coherent string states. We derive the expression
for the norm of such states and show that the condition of finite positive
norm leads to a mass-spectrum as explained above. We also show that our
coherent string states lead to non-vanishing expectation values only for the
components of the currents corresponding to the classical
oscillating circular strings.
Finally in Section 6, we have some concluding remarks.
\section{SL(2,R) WZWN Model. The Classical Picture.}
\setcounter{equation}{0}
Our starting point is the sigma-model action including the WZWN term
at
level
$k$ \cite{wit}
\begin{equation}
S=-\frac{k}{8\pi}\int_{M} d\tau
d\sigma\;\eta^{\alpha\beta}\mbox{Tr}[
g^{-1}\partial_\alpha g\;g^{-1}\partial_\beta g]-
\frac{k}{12\pi}\int_{B} \mbox{Tr}[
g^{-1}dg\wedge g^{-1}dg\wedge g^{-1}dg]
\end{equation}
Here $M$ is the boundary of the manifold $B$, and $g$ is a
group-element of the group under consideration (later taken to be
$SL(2,R)$).
The classical string equations of motion are
\begin{equation}
\partial_{-}(g^{-1}\partial_{+}g)=0
\end{equation}
where we introduced world-sheet light-cone coordinates $\sigma^{\pm}=\tau\pm
\sigma$. The world-sheet energy-momentum tensor is
\begin{equation}
T_{\pm\pm}=-\frac{2}{k}{\mbox{Tr}}(J_\pm J_\pm)
\end{equation}
where the conserved currents, $\partial_\pm J_\mp=0$, are given by
\begin{equation}
J_+=ikg^{-1}(\partial_+ g),\;\;\;\;\;\;J_-=-ik(\partial_-g)g^{-1}
\end{equation}
and the string constraints are
\begin{equation}
{\mbox{Tr}}[(g^{-1}\partial_\pm g)(g^{-1}\partial_\pm g)]=0
\end{equation}
Equation (2.2) is trivially solved by \cite{wit}
\begin{equation}
g(\sigma^+,\sigma^-)=g_R(\sigma^-)g_L(\sigma^+)
\end{equation}
It follows that
\begin{equation}
J_+=ikg_L^{-1}(\partial_+ g_L),\;\;\;\;\;J_-=-ik(\partial_- g_R)g_R^{-1}
\end{equation}
and the constraints, eq.(2.5), separate
\begin{equation}
{\mbox{Tr}}[(g_L^{-1}\partial_+ g_L)^2]={\mbox{Tr}}
[(g_R^{-1}\partial_- g_R)^2]=0
\end{equation}
In the case of $SL(2,R)$, the group elements are given by
\begin{equation}
g_L(\sigma^+)=\left( \begin{array}{cc} \tilde{a}(\sigma^+) &
\tilde{u}(\sigma^+) \\
-\tilde{v}(\sigma^+) & \tilde{b}(\sigma^+) \end{array}\right), \;\;\;\;\;\;\;
\;\; g_R(\sigma^-)=\left( \begin{array}{cc} a(\sigma^-) & u(\sigma^-) \\
-v(\sigma^-) & b(\sigma^-) \end{array}\right)
\end{equation}
subject to the normalization conditions
\begin{equation}
\tilde{a}(\sigma^+)\tilde{b}(\sigma^+)+
\tilde{u}(\sigma^+)\tilde{v}(\sigma^+)=a(\sigma^-)b(\sigma^-)+
u(\sigma^-)v(\sigma^-)=1
\end{equation}
Then the constraints, eqs.(2.8), are simply
(from now on we do not write explicitly the arguments $(\sigma^\pm)$
of the functions)
\begin{equation}
\tilde{a}_+\tilde{b}_+ + \tilde{u}_+\tilde{v}_+=a_- b_- + u_- v_- =0
\end{equation}
where we introduced the notation $a_-=\partial_- a$, $\tilde{a}_+=
\partial_+\tilde{a}$, etc.

As for the currents, it is convenient to make a Pauli decomposition
\begin{equation}
J_\pm=\eta_{ab}J^a_\pm t^b
\end{equation}
in terms of the Pauli matrices,
\begin{equation}
t^1=\frac{i}{2}\sigma^1,\;\;\;\;\; t^2=-\frac{i}{2}\sigma^3,\;\;\;\;\; t^3=
\frac{1}{2}\sigma^2
\end{equation}
such that
\begin{equation}
{\mbox{Tr}}(t^a t^b)=-\frac{1}{2}\eta^{ab},\;\;\;\;\;\;\;\;
[t^a,t^b]=i\epsilon^{abc}t_c
\end{equation}
(Our conventions are: $\eta^{ab}={\mbox{diag}}(1,1,-1)$ and
$\epsilon^{123}=+1$).

It is also standard to introduce
\begin{equation}
J_-^\pm=J_-^1\pm iJ_-^2,\;\;\;\;\;\;\;\;J_+^\pm=J_+^1\pm iJ_+^2
\end{equation}
It is now straightforward to write down explicit expressions for the currents
in terms of the group elements, eqs.(2.9),
\begin{eqnarray}
J_-^\pm&=&-k\left( [au_- -ua_- +vb_- -bv_-]\pm 2i[ab_- +uv_-]\right)
\nonumber\\
J_-^3&=&k[vb_- -bv_- +ua_- -au_-]\nonumber\\
J_+^\pm&=&k\left( [\tilde{b}\tilde{u}_+ -\tilde{u}\tilde{b}_+ +
\tilde{v}\tilde{a}_+ -\tilde{a}\tilde{v}_+]\pm 2i
[\tilde{v}\tilde{u}_+ +\tilde{a}\tilde{b}_+]\right)\nonumber\\
J_+^3&=&-k[\tilde{v}\tilde{a}_+ -\tilde{a}\tilde{v}_+ +\tilde{u}\tilde{b}_+ -
\tilde{b}\tilde{u}_+]
\end{eqnarray}
Notice also that
\begin{equation}
T_{\pm\pm}=\frac{1}{k}(J_\pm^+ J_\pm^- - J_\pm^3 J_\pm^3)
\end{equation}
such that the conditions $T_{\pm\pm}=0$ again lead to eqs.(2.11), as
they should.

We close this section with a few comments about the invariant string size
and the reduction of the classical equations of motion
to the Liouville equation \cite{liou} (for a review of different
methods, see Ref .\cite{raif}).

The line-element on the group manifold is given by
\begin{equation}
dS^2=\frac{1}{H^2}{\mbox{Tr}}[(g^{-1}dg)^2]
\end{equation}
where $H^{-1}$ is the length-scale, which up to a numerical factor
is related to $k$ and $\alpha'$ by  \cite{wit,yan}
\begin{equation}
k=\frac{1}{H^2\alpha'}
\end{equation}
where $\alpha'$ is related to the string tension $T$ in the usual way,
$T=(2\pi\alpha')^{-1}$.

The line-element on the group manifold
induces the following proper line-element on the string world-sheet
\begin{equation}
ds^2=-\frac{e^\alpha}{2H^2}\;d\sigma^+
d\sigma^-
\end{equation}
Here $\alpha=\alpha(\sigma^+,\sigma^-)$ is the
fundamental quadratic form, which determines the invariant string size, and is
defined by
\begin{equation}
e^\alpha\equiv -{\mbox{Tr}}[(g^{-1}\partial_+ g)(g^{-1}\partial_- g)]
\end{equation}
In the case of $SL(2,R)$, one finds
\begin{eqnarray}
e^\alpha&=&[av_- -va_-][\tilde{a}\tilde{u}_+ -\tilde{u}\tilde{a}_+]+
[ub_- -bu_-][\tilde{v}\tilde{b}_+ -\tilde{b}\tilde{v}_+]\nonumber\\
&+&2[ab_- +vu_-][\tilde{b}\tilde{a}_+ +\tilde{v}\tilde{u}_+]
\end{eqnarray}
By differentiating this identity twice and by using some simple algebra,
we get the equation
\begin{equation}
\alpha_{+-}=2f(\sigma^-)\tilde{g}(\sigma^+)e^{-\alpha}
\end{equation}
where the functions $f=f(\sigma^-)$ and $\tilde{g}=\tilde{g}(\sigma^+)$ are
given by
\begin{eqnarray}
f&=&\frac{1}{2}\left( \frac{u_-}{a_-}[av_{--}-va_{--}]-
\frac{v_-}{b_-}[ub_{--}-bu_{--}]\right)\\
\tilde{g}&=&\frac{1}{2}\left( \frac{\tilde{u}_+}{\tilde{b}_+}
[\tilde{v}\tilde{b}_{++}-\tilde{b}\tilde{v}_{++}]-
\frac{\tilde{v}_+}{\tilde{a}_+}[\tilde{a}\tilde{u}_{++}-
\tilde{u}\tilde{a}_{++}]\right)
\end{eqnarray}
The product $f(\sigma^-)\tilde{g}(\sigma^+)$ in eq.(2.23) can be absorbed by a
conformal transformation, thus we conclude that the most
general
equation fulfilled by the fundamental quadratic form
$\alpha$
is
\begin{equation}
\alpha_{+-}+
Ke^{-\alpha}=0,
\end{equation}
where:
\begin{equation}
K=\left\{ \begin{array}{l}
+1,\;\;\;\;\;\;f(\sigma^-) \tilde{g}(\sigma^+)<0 \\
-1,\;\;\;\;\;\;f(\sigma^-) \tilde{g}(\sigma^+)>0 \\
\;0,\;\;\;\;\;\;\;\;f(\sigma^-)
\tilde{g}(\sigma^+)=0
\end{array}\right.
\end{equation}
Equation (2.26) is either the
Liouville equation ($K=\pm 1$), or the free wave equation $(K=0)$.
This result was obtained in a different way and discussed in detail in Ref.
\cite{liou}.
\section{Circular Strings}
\setcounter{equation}{0}
Circular strings on the $SL(2,R)$ group manifold were considered in detail
in Ref.
\cite{vega}. In this section we translate the results into the formalism of
$SL(2,R)$ currents.

Circular strings are most easily discussed using a
different parametrization of $SL(2,R)$, corresponding to the static
coordinates
for Anti de Sitter spacetime.
We first write the $SL(2,R)$ group-element in the form
\begin{equation}
g=g_R g_L=\left( \begin{array}{cc} A & U \\
-V & B \end{array} \right)=\left( \begin{array}{cc} a\tilde{a}-u\tilde{v}
& a\tilde{u}+u\tilde{b} \\
-v\tilde{a}-b\tilde{v} & b\tilde{b}-v\tilde{u} \end{array} \right)
\end{equation}
and then introduce coordinates $(t,r,\phi)$ by
\begin{eqnarray}
A&=&\sqrt{1+H^2r^2}\;\cos(Ht)+Hr\cos(\phi)\nonumber\\
B&=&\sqrt{1+H^2r^2}\;\cos(Ht)-Hr\cos(\phi)\nonumber\\
U&=&\sqrt{1+H^2r^2}\;\sin(Ht)-Hr\sin(\phi)\nonumber\\
V&=&\sqrt{1+H^2r^2}\;\sin(Ht)+Hr\sin(\phi)
\end{eqnarray}
In these coordinates, the line-element (2.18) on the group manifold becomes
\begin{equation}
dS^2=-(1+H^2r^2)dt^2+\frac{dr^2}{1+H^2r^2}+r^2d\phi^2
\end{equation}
i.e., the standard parametrization of $2+1$ dimensional Anti de Sitter
spacetime using static coordinates \cite{rind}. As usual we unwind the
temporal
coordinate $t$, corresponding to
considering the covering group of $SL(2,R)$.
Moreover, the anti-symmetric
tensor which can be read off from eq.(2.1), is given
by
\begin{equation}
B_{t\phi}=-B_{\phi t}=\frac{1}{2}Hr^2\end{equation}
with all other components vanishing.

In the $(t,r,\phi)$ coordinates, the oscillating
circular strings are given by \cite{vega}
\begin{eqnarray}
\phi&=&\sigma\nonumber\\
Ht&=&\arctan\left(
\frac{1+EH}{\sqrt{1+2EH}}\tan(\sqrt{1+2EH}\;\tau)\right)
-\tau\nonumber\\
r&=&\frac{E}{\sqrt{1+2EH}} \sin(\sqrt{1+2EH}\;\tau)
\end{eqnarray}
where $E$ is an integration constant.

It is now straightforward to work backwards and read off the explicit
expressions for the leftmoving and rightmoving group-elements, eqs.(2.9). The
expressions are however not very enlightening, so we give them in Appendix A.
It is more interesting to consider directly the leftmoving and rightmoving
currents, eqs.(2.16). After some simple algebra, they are found to be
\begin{eqnarray}
J_-^\pm&=&\pm iEHke^{\pm i\sigma^-}\nonumber\\
J_-^3&=&-EHk\nonumber\\
J_+^\pm&=&\mp iEHk e^{\mp i\sigma^+}\nonumber\\
J_+^3&=&EHk
\end{eqnarray}
Thus, the circular strings contain only modes corresponding to $n=0$ and
$n=\pm 1$. This was of course to be expected for a circular string, c.f.
circular strings in Minkowski spacetime, but it is in fact highly
implicit in the parametrization (3.5). From the conformal field theory
point of view, the parametrization (3.6) is thus more natural.
\section{Quantization}
\setcounter{equation}{0}
In this section we give a short review of quantization of conformal field
theories, corresponding to bosonic strings on
group manifolds (see for instance Refs. \cite{wit2,evans}).
This is mainly to fix our
conventions and normalizations. We only give the results which we
shall use in Section 5.

The currents $J^a_\pm$, as introduced in eq.(2.12), can be expanded in
Fourier series
\begin{equation}
J^a_-=\sum_{n=-\infty}^{\infty}\;J^a_n\;e^{-in\sigma^-};\;\;\;\;\;\;
\;\;\left( J^a_n\right) ^\dagger=J^a_{-n}
\end{equation}
and similarly for $J^a_+$ in terms of $\sigma^+$.
In the following we shall  consider only the
rightmovers $(-)$; the construction for the leftmovers $(+)$
is of course similar. For
simplicity we shall therefore also skip the minus indices
of $J^a_-$, $T_{--}$ etc.
The $SL(2,R)$ Kac-Moody algebra is
\begin{equation}
[J^a_m,J^b_n]=i\epsilon^{ab}\;_c J^c_{m+n}+\frac{k}{2}m\eta^{ab}\delta_{n+m}
\end{equation}
In terms of the currents, eq.(2.15), the algebra becomes
\begin{eqnarray}
&[J^+_m,J^-_n]&= -2J^3_{m+n}+km\delta_{m+n}\nonumber\\
&[J^3_m,J^{\pm}_n]&=\pm J^{\pm}_{m+n}\nonumber\\
&[J^3_m,J^3_n]&=-\frac{k}{2}m\delta_{m+n}
\end{eqnarray}
At the quantum level, the world-sheet energy-momentum tensor takes the
Sugawara form
\begin{equation}
T=\frac{1}{k-2}\;\eta_{ab}:J^a J^b:\;=\frac{1}{k-2}
:\left(J^+J^- -J^3 J^3\right):
\end{equation}
Its Fourier modes
\begin{equation}
T=\sum_{n=-\infty}^\infty\;L_n\;e^{-in\sigma^-}
\end{equation}
are given by
\begin{equation}
L_n=\frac{1}{k-2}\sum_{l=-\infty}^\infty\; :\left( \frac{1}{2}(J^+_{n-l}J^-_l+
J^-_{n-l}J^+_l)-J^3_{n-l}J^3_l\right):
\end{equation}
They fulfill the Virasoro algebra
\begin{equation}
[L_m,L_n]=(m-n)L_{m+n}+\frac{c}{12}m(m^2-1)\delta_{m+n}
\end{equation}
where the central charge is given by
\begin{equation}
c=\frac{3k}{k-2}
\end{equation}
Demanding $c=26$, corresponding to conformal invariance, gives $k=52/23$.
Notice also the commutators
\begin{equation}
[L_n,J^{\pm}_m]=-mJ^{\pm}_{n+m},\;\;\;\;\;\;\;\;
[L_n,J^3_m]=-mJ^3_{n+m}
\end{equation}
which will be usefull in the following.

The Kac-Moody algebra contains the subalgebra of zero modes $J^a_0$, for which
the quadratic Casimir is
\begin{equation}
Q=\eta_{ab}J^a_0 J^b_0=\frac{1}{2}\left( J^+_0 J^-_0 +J^-_0 J^+_0\right)-
J^3_0 J^3_0
\end{equation}
The primary states, which are quantum
states $|jm>$ at grade zero ("base-states" or "ground-states"), are
characterised by
\begin{equation}
Q|jm>=-j(j+1)|jm>,\;\;\;\;\;\;\;\;J^3_0|jm>=m|jm>
\end{equation}
Moreover, they fulfill
\begin{equation}
J^{\pm}_0|jm>=\sqrt{m(m\pm 1)-j(j+1)}\;|jm\pm 1>
\end{equation}
as well as
\begin{equation}
J^a_l|jm>=0;\;\;\;\;\;l>0
\end{equation}
The primary
states must belong to one of the unitary representations of $SL(2,R)$
(or its covering group) \cite{barg,dixon}.
We shall return to this point in the next section.

From the primary states, one can construct the excited states
as descendents by applying
$J^a_{-l}$ operators ($l$ is a positive integer)
\begin{equation}
|\psi>=J^{a_1}_{-l_1}J^{a_2}_{-l_2}.....J^{a_r}_{-l_r}|jm>
\end{equation}
and so on. The physical state conditions (the mass-shell condition and the
Virasoro primary condition) are then
\begin{equation}
(L_0-1)|\psi>=0
\end{equation}
\begin{equation}
L_l|\psi>=0;\;\;\;\;\;l>0
\end{equation}
For a physical state of the form (4.14) at grade $n=\sum l_{i}$, the mass-shell
condition gives
\begin{equation}
n-\frac{j(j+1)}{k-2}=1
\end{equation}
Identifying (minus) the quadratic Casimir, eq.(4.10), of
the base-state with the mass-squared \cite{nem}, or more precisely we
normalize the mass as
\begin{equation}
m^2\alpha'\equiv\frac{j(j+1)}{k-2}
\end{equation}
we then see that the mass-squared grows
linearly with the grade
\begin{equation}
m^2\alpha'=n-1
\end{equation}
This is just like for strings
in flat Minkowski spacetime.
\section{Coherent String States}
\setcounter{equation}{0}
The idea is now to construct exact quantum states with properties similar to
the classical circular strings considered in Section 3. More precisely, our aim
is to construct states $|\psi>$ such that the only components of the
currents giving non-vanishing expectation values, $<\psi|J^a|\psi>\neq 0$,
are those components corresponding to the non-vanishing classical currents,
eq.(3.6).
In other words, considering for simplicity only the rightmovers, then only the
components $J^+_{-1}$, $J^-_{+1}$ and $J^3_0$ should have non-vanishing
expectation values; all other components $J^+_n\;(n\neq -1)$,
$J^-_n\;(n\neq+1)$ and $J^3_n\;(n\neq 0)$ must have zero expectation values.

Several problems immediately appear: If we consider states of the form
(4.14), it would be
possible to obtain a non-vanishing expectation value for $J^3_0$,
but it would certainly be impossible to get non-vanishing expectation values of
$J^+_{-1}$ and $J^-_{+1}$. Moreover, as mentioned at the end of Section 4,
states of the form (4.14) give rise to a mass-spectrum where the
mass-squared grows linearly with the grade as in flat space.
However, semi-classical
quantization of the circular strings has been shown \cite{vega} to lead to a
mass-spectrum where $m^2\alpha'\propto N^2$ ($N$ positive
integer), at least for the high
mass states.
Fortunately, it turns out that both problems can be solved by considering
{\it coherent}
string states on the $SL(2,R)$ group manifold (a similar construction
for the $SU(2)$ group manifold was considered in \cite{niels}).

As a starting point, we consider states of the form
\begin{equation}
\left( J^+_{-1}\right) ^n|jj>\; ;\;\;\;\;\;\;\;\;n\geq 0
\end{equation}
where $|jj>$ belongs to the highest weight discrete series $D^-_j$
\cite{barg,dixon}, with states $|jm>$
\begin{equation}
j\leq -1/2\;,\;\;\;\;\;\;\;\;m=j,j-1,...
\end{equation}
Since we shall consider the covering group
of $SL(2,R)$, there are no further restrictions on $j$, i.e., it need not be
integer or half-integer \cite{barg,dixon}.
In particular, from eq.(4.12) it follows that
\begin{equation}
J^+_0|jj>=0\; ,\;\;\;\;\;\;\;\; J^-_0|jj>=\sqrt{-2j}\;|jj-1>
\end{equation}
The states eq.(5.1) fulfill the Virasoro primary condition
\begin{equation}
L_l\left( J^+_{-1}\right)|jj>=0;\;\;\;\;\;\;\;\;l>0
\end{equation}
and the mass-shell condition (4.15) leads to
\begin{equation}
n=1+\frac{j(j+1)}{k-2}\;\;\Leftrightarrow\;\;j=-\frac{1}{2}-
\sqrt{(k-2)(n-1)+1/4}
\end{equation}
However, these states generally do not have positive norm. Indeed
\begin{equation}
<jj|\left( J^-_{+1}\right)^m\left( J^+_{-1}\right)^n|jj>=
n!\;\delta_{nm}\;\prod_{i=1}^{n}\left( k-2+i-\sqrt{4(k-2)(n-1)+1}\;\right)
\end{equation}
and the right hand side is generally not positive. For example, it is negative
for $n=m=2$ and $n=m=3$ (using that $k=52/23$). This is of course just a
simple
example illustrating the well known unitarity problem for strings on $SL(2,R)$
[1-10, 13] (for recent reviews, see
\cite{petr1,bars2}).

We consider instead coherent states built from states of the form (5.1)
\begin{equation}
e^{\mu J^+_{-1}}|jj>
\end{equation}
where $\mu$ is an arbitrary
complex number. These states certainly also fulfill the
Virasoro
primary condition but, being coherent states, they obviously are eigenstates
of neither the number operator nor of the $L_0$ operator.
We shall therefore impose a "weak" mass-shell condition
\begin{equation}
<jj|e^{\mu^* J^-_{+1}}\left( L_0 -1\right) e^{\mu J^+_{-1}}|jj>=0
\end{equation}
Before evaluating the left hand side of eq.(5.8), it is necessary
to consider the normalization of the states (5.7).

In ordinary quantum mechanics with creation and annihilation operators
$a^\dagger$ and $a$, respectively, and a vacuum state $|0>$
\begin{equation}
[a,a^\dagger]=1\; ,\;\;\;\;\;\;\;\; a|0>=0
\end{equation}
the excited states are constructed as
\begin{equation}
\left( a^\dagger \right)^n |0>=\sqrt{n!}|n>\; , \;\;\;\;\;\;\;\;<n|m>=
\delta_{nm}
\end{equation}
In that case, a coherent state can always be normalized. In fact, the coherent
state
\begin{equation}
|\mu>\equiv e^{-\frac{1}{2}\mu^*\mu}\;e^{\mu a^\dagger}|0>
\end{equation}
has unit norm, for arbitrary complex number $\mu$. Notice also that
the coherent state is an eigenstate of the annihilation
operator
\begin{equation}
a|\mu>=\mu|\mu>
\end{equation}
which can be taken as the definition of coherent states in ordinary
quantum mechanics.
(For more discussion, see for instance
\cite{merz}).

In our case, the situation is somewhat different since we have a Kac-Moody
algebra (4.2) with a non-Abelian term in the current algebra,
and in particular since the group manifold $SL(2,R)$ is non-compact and has a
time-like direction (contrary to the case of $SU(2)$ \cite{niels}).
It implies that the coherent state (5.7) is not an eigenstate of the
annihilation operator $J^-_{+1}$
\begin{equation}
J^-_{+1} e^{\mu J^+_{-1}}|jj>=\mu\left( 2j+k+\mu J^+_{-1}\right)
e^{\mu J^+_{-1}}|jj>
\end{equation}
Moreover, the coherent state (5.7) can not be normalized for arbitrary
complex number $\mu$.
One finds
\begin{equation}
<jj|e^{\mu^* J^-_{+1}}\; e^{\mu J^+_{-1}}|jj>=1+
\sum_{n=1}^\infty\frac{(\mu^*\mu)^n}{n!}\prod_{l=1}^n (2j+k-1+l)
\end{equation}
The product on the right hand side goes as $n!$. Thus the infinite sum is
convergent
only if $\mu^*\mu<1$, or if the infinite sum terminates after a finite
number of terms (this happens if $2j+k-1+l=0$, for some $l$).
More precisely, the right hand side of eq.(5.14) is a finite positive
number in the following two cases\\
\\
{\bf (I):} $\mu^*\mu<1$ and $j$ arbitrary ($j\leq -1/2$).\\
In this case the normalized state is
\begin{equation}
|\mu I>=(1-\mu^*\mu)^{j+k/2}\;e^{\mu J^+_{-1}}|jj>
\end{equation}
\\
{\bf (II):} $\mu^*\mu>1$ and $j=-N-k/2\;\;\;\;(N={0,1,2,...}).$\\
In this case the normalized state is
\begin{equation}
|\mu II>=(\mu^*\mu-1)^{-N}\;e^{\mu J^+_{-1}}|-N-k/2, -N-k/2>
\end{equation}
Let us now return to the mass-shell condition, eq.(5.8), which
gives rise to some additional
constraints on $\mu$ and $j$. In the two cases, respectively, one finds\\
\\
{\bf (I):}\\
\begin{equation}
\mu^*\mu=\frac{1+\frac{j(j+1)}{k-2}}{2j+k+1+\frac{j(j+1)}{k-2}}<1\; ;\;\;\;\;
\;\;\;\;-\frac{k}{2}<j\leq -\frac{1}{2}
\end{equation}\\
\\
{\bf (II):}\\
\begin{equation}
\mu^*\mu=\frac{1+\frac{j(j+1)}{k-2}}{2j+k+1+\frac{j(j+1)}{k-2}}>1\; ;\;\;\;\;
\;\;\;\; j=-N-\frac{k}{2}\;\;\;\;(N={1,2,...})
\end{equation}
Thus, the spectrum consists of two parts: {\bf (I)}
A continuous spectrum where $j$
fulfills the standard spin-level condition
[2-4, 6-10, 13, 14]
$-k/2 <j\leq -1/2$, and {\bf (II)}
a discrete spectrum where $j$ fulfills  $j=-N-k/2\;$ ($N$ positive
integer).
If we were considering ordinary descendent states of the form (4.14), we
would generally
not be allowed to have $|j|>k/2$ because of unitarity, thus the discrete part
{\bf (II)} would not be allowed. For the coherent
states under consideration here, there is however no problem since the
quantization condition $j=-N-k/2$ ($N$ positive integer) precisely
ensures that they
are all positive norm states.

Introducing the mass with the normalization as in eq.(4.18), we find that
the continuous part of the spectrum {\bf (I)} corresponds to
\begin{equation}
m^2\alpha'\in\; [\frac{-1}{4(k-2)},\;\frac{k}{4}[\;\;=\;[-\frac{23}{24},\;
\frac{13}{23}[
\end{equation}
where the last equality was obtained using $k=52/23$, corresponding to
conformal
invariance. That is, the continuous part of the spectrum {\bf (I)} consists
of low
mass states and is partly tachyonic.

On the other hand, the discrete spectrum {\bf (II)} gives
\begin{equation}
m^2\alpha'=\frac{N^2}{k-2}\left( 1+\frac{k-1}{N}+\frac{k(k-2)}{4N^2}\right)
\end{equation}
i.e., for the discrete part of the spectrum we find that
$m^2\alpha'\propto N^2$ ($N$
positive integer). Asymptotically,
this is precisely what was found using semi-classical
quantization \cite{san2,vega1,vega}.
It should be stressed, however, that $N$ is not the
eigenvalue of the number operator; as already mentioned, since we are
working with coherent states, we do not have any eigenstates of the
number operator. Thus $N$ is simply a positive integer here.

Notice also that $k$, $\alpha'$ and  the length-scale $H$ in the
quantum theory are related as in
eq.(2.19), but with $k$ replaced by $k-2$.
With the present normalizations we therefore have  exactly the same
leading order behaviour, including the numerical
coefficient, for the mass-squared
as obtained using semi-classical quantization in Ref. \cite{vega}.

Finally, the relation with the classical circular strings is established
by noting that only the expectation values of $J^+_{-1}$, $J^-_{+1}$ and
$J^3_0$ are non-vanishing ($i=I,II$)
\begin{eqnarray}
<\mu i|J^+_l|\mu i>&=&\left\{ \begin{array}{cc}
(2j+k)\frac{\mu^*}{1-\mu^*\mu}\;
,\;\;\;\;\;\;l=-1 \\
0\; ,\;\;\;\;\;\; l\neq -1 \end{array}\right.\nonumber\\
<\mu i| J^-_l|\mu i>&=&\left\{ \begin{array}{cc} (2j+k)\frac{\mu}{1-\mu^*\mu}\;
,\;\;\;\;\;\; l=+1 \\
0\; ,\;\;\;\;\;\; l\neq +1 \end{array}\right. \\
<\mu i| J^3_l |\mu i>&=&\left\{ \begin{array}{cc} j+(2j+k)\frac{\mu^*\mu}
{1-\mu^*\mu}\; ,\;\;\;\;\;\; l=0 \\
0\; , \;\;\;\;\;\;l\neq 0 \end{array}\right. \nonumber
\end{eqnarray}
valid for both $|\mu I>$ and $|\mu II>$ for the respective values of
$\mu$ and $j$, as given in eqs.(5.17)-(5.18).
To obtain these results, as well as most other results in this section, we used
the commutators listed in Appendix B.
\section{Conclusion}
\setcounter{equation}{0}
We have shown that very massive string states in the $SL(2,R)$ WZWN model
(corresponding to $AdS_3$) can be described as {\it coherent}
states based on the
standard discrete representation $D_j^-$. The spectrum of such states
was shown to consist of two parts: A continuous low mass (partly tachyonic)
part and a discrete high mass part. For the continuous part, the spin
$j$ must fulfill the standard spin-level restriction $-k/2<j\leq -1/2$,
while for the discrete part we get the quantization condition
$j=-N-k/2$ ($N$ positive integer). Although the latter seems to be in
contradiction with the spin-level restriction, the quantization condition
precisely ensures that {\it all} our coherent states have finite positive
norm. Thus, no ghost-states are included in the spectrum.
The mass spectrum of the discrete part of the spectrum, eq.(5.20), shows
the asymptotic behaviour $m^2\alpha'\propto N^2$. This is in precise
agreement with our previous results obtained using semi-classical
quantization \cite{san2,vega1,vega}.
The same asymptotic behaviour was also obtained in
the recent preprint \cite{Ooguri}, although the construction there is
completely different from ours, as discussed in more detail in the
introduction.

In this paper we used, for simplicity and clarity, the example of an
oscillating circular string. That is, our quantum coherent string states
were constructed to lead to non-vanishing  expectation values for
very specific components of the currents, eq.(5.21). It is however
easy to generalize our construction to other string configurations too.

We saw in Section 5 that the coherent states, eq.(5.7), do not have all the
same
properties of standard quantum mechanical coherent states \cite{merz}. For
instance, they are not eigenstates of the annihilation operator.  All the
standard properties are however obtained in the following manner: First we
must renormalize the currents $J^a_n\rightarrow \sqrt{k}J^a_n$, as follows
from the algebra, eq.(4.3) (although
this does not work for the zero-modes \cite{nem}). Secondly,
for the coherent
states we must renormalize
the complex number $\mu$ by $\mu\rightarrow \mu/\sqrt{k}$,
as follows from eq.(5.7). Finally, we let $k\rightarrow \infty$. Then, it
follows that the non-Abelian piece drops out in eq.(5.13) and we get
an eigenstate of the renormalized annihilation operator. More generally,
by the same prescription we recover all the usual well-known properties of
standard quantum mechanical coherent states \cite{merz}. For instance,
the continuous spectrum, represented by the states (5.15), will now be
valid for any $j\leq -1/2$, and the states become
\begin{equation}
|\mu I> \rightarrow e^{-\frac{1}{2}\mu^*\mu}\;e^{\mu J^+_{-1}} |jj>;
\;\;\;\;\;\;\;\; k\rightarrow\infty
\end{equation}
c.f. eq.(5.11). On the other hand,
the discrete part of the spectrum, represented by the
states (5.16), disappears since all the states  are pushed to infinite mass.
These are precisely the properties which characterise the usual coherent
states of quantum mechanics as {\it quasi-classical} states, for
which the quantum uncertainty is minimal.
\setcounter{equation}{0}
\vskip 12pt
{\bf Acknowledgements}\\
A.L.L. would like to thank the {\bf Ambassade de France \`a Copenhague},
{\bf Service Culturel et Scientifique}  for financial support in Paris,
during the preparation of this paper.
\section{Appendix A}
\setcounter{equation}{0}
In this appendix we give the explicit expressions for the group-elements (2.9)
corresponding to the circular strings (3.5). For simplicity, we only give
the results for the rightmovers.
\begin{eqnarray}
a(\sigma^-)&=&\frac{1}{\sqrt{1+2EH}}\sin\left( \sqrt{1+2EH}\;\frac{\sigma^-}{2}
\right)\left[ (1+EH)\sin\left(\frac{\sigma^-}{2}\right)+EH\cos\left(
\frac{\sigma^-}{2}\right)\right]\nonumber\\
&+&\cos\left(\frac{\sigma^-}{2}\right)\cos\left(\sqrt{1+2EH}\;\frac{\sigma^-}{2}
\right)
\end{eqnarray}
\begin{eqnarray}
b(\sigma^-)&=&\frac{1}{\sqrt{1+2EH}}\sin\left( \sqrt{1+2EH}\;\frac{\sigma^-}{2}
\right)\left[ (1+EH)\sin\left(\frac{\sigma^-}{2}\right)-EH\cos\left(
\frac{\sigma^-}{2}\right)\right]\nonumber\\
&+&\cos\left(\frac{\sigma^-}{2}\right)\cos\left(\sqrt{1+2EH}\;\frac{\sigma^-}{2}
\right)
\end{eqnarray}
\begin{eqnarray}
u(\sigma^-)&=&\frac{1}{\sqrt{1+2EH}}\sin\left(\sqrt{1+2EH}\;\frac{\sigma^-}{2}
\right)\left[ (1+EH)\cos\left(\frac{\sigma^-}{2}\right)+EH\sin\left(
\frac{\sigma^-}{2}\right)\right]\nonumber\\
&-&\sin\left(\frac{\sigma^-}{2}\right)\cos\left(\sqrt{1+2EH}\;\frac{\sigma^-}{2}
\right)
\end{eqnarray}
\begin{eqnarray}
v(\sigma^-)&=&\frac{1}{\sqrt{1+2EH}}\sin\left( \sqrt{1+2EH}\;\frac{\sigma^-}{2}
\right)\left[ (1+EH)\cos\left(\frac{\sigma^-}{2}\right)-EH\sin\left(
\frac{\sigma^-}{2}
\right)\right]\nonumber\\
&-&\sin\left(\frac{\sigma^-}{2}\right)\cos\left(\sqrt{1+2EH}\;\frac{\sigma^-}{2}
\right)
\end{eqnarray}
Now using eq.(2.16), it is straightforward to obtain (3.6) for the rightmovers.
The derivation for the leftmovers is similar.
\section{Appendix B}
\setcounter{equation}{0}
In this appendix we list some useful commutators used in Section 5.
\begin{equation}
[J^3_m,\left( J^+_{-1}\right)^n]=nJ^+_{m-1}\left(J^+_{-1}\right)^{n-1}
\end{equation}
\begin{equation}
[J^+_m,\left(J^-_{+1}\right)^n]=-n\left(2J^3_{m+1}-km\delta_{m+1}\right)
\left(J^-_{+1}\right)^{n-1}-n(n-1)J^-_{m+2}\left(J^-_{+1}\right)^{n-2}
\end{equation}
\begin{equation}
[J^-_m,\left(J^+_{-1}\right)^n]=n\left(J^+_{-1}\right)^{n-1}\left(2J^3_{m-1}
+km\delta_{m-1}\right)+n(n-1)\left(J^+_{-1}\right)^{n-2}J^+_{m-2}
\end{equation}
These identities are most easily proved by induction using eqs.(4.3).
It follows that
\begin{equation}
[J^3_m,e^{\mu J^+_{-1}}]=\mu J^+_{m-1} e^{\mu J^+_{-1}}
\end{equation}
\begin{equation}
[J^+_m,e^{\mu^* J^-_{+1}}]=-\mu^*\left( 2J^3_{m+1}-km\delta_{m+1}
+\mu^* J^-_{m+2}\right) e^{\mu^* J^-_{+1}}
\end{equation}
\begin{equation}
[J^-_m,e^{\mu J^+_{-1}}]=\mu e^{\mu J^+_{-1}}\left( 2J^3_{m-1}+km\delta_{m-1}
+\mu J^+_{m-2}\right)
\end{equation}

\end{document}